\documentclass{PoS}

\usepackage{latexsym}
\usepackage{epstopdf}
\usepackage{epsfig}
\usepackage{multicol}
\usepackage{pstricks,pst-grad}
\usepackage{cite}

\usepackage{amsmath}
\usepackage{amsfonts}
\usepackage{amssymb}
\usepackage{rotating}

\newcommand{\be}{ \begin{equation} }
\newcommand{\ee}{\end{equation}}
\newcommand{\bea}{\begin{eqnarray}}
\newcommand{\eea}{\end{eqnarray}}
\newcommand{\bit}{\begin{itemize}}
\newcommand{\eit}{\end{itemize}}
\newcommand{\bmp}[1]{\begin{minipage}{#1cm}}
\newcommand{\emp}{\end{minipage}}

\newcommand{\Tr}{\mbox{Tr}}

\newcommand{\SL}{$\rm{SL}(3,\mathbb{C})$}
\newcommand{\SU}{SU($N$)\,}

\newcommand{\bra}{\langle}
\newcommand{\ket}{\rangle}

\title{Complex Langevin dynamics for SU(3) gauge theory in the presence of
a theta term}
\ShortTitle{Complex Langevin dynamics in the presence of
a theta term}

\author{\speaker{Lorenzo Bongiovanni} \\
Department of Physics, College of Science, Swansea University, Swansea, United Kingdom\\
E-mail: \email{pylb@pyserver.swan.ac.uk}}
\author{Gert Aarts\\
Department of Physics, College of Science, Swansea University, Swansea, United Kingdom\\
E-mail: \email{G.Aarts@swansea.ac.uk}}
\author{Erhard Seiler\\
Max-Planck-Institut f\"ur Physik (Werner-Heisenberg-Institut), M\"unchen, Germany\\
E-mail: \email{ehs@mpp.mpg.de}}
\author{D\'enes Sexty\\
Institut f\"ur Theoretische Physik, Universit\"at Heidelberg, Heidelberg, Germany\\
E-mail: \email{D.Sexty@ThPhys.Uni-Heidelberg.DE}}

\abstract{
One of the yet unsolved questions of QCD in the context of the Standard Model is to explain the strong CP problem.  A way to look for a better understanding of it is to investigate the theory in the presence of a non-zero topological $\theta$-term. On the lattice such a term is complex: hence it introduces a sign problem which, in general, limits the applicability of standard Monte Carlo methods.  Here we will discuss the approach of complex Langevin dynamics and show results for both real and imaginary values of $\theta$. We also report on our experience with the gradient flow for real and imaginary $\theta$.   
}

\FullConference{The 32nd International Symposium on Lattice Field Theory\\
23-28 June, 2014\\
Columbia University New York, NY}

\begin{document}

\section{Introduction}

Non-perturbative effects play a fundamental role in many aspects of QCD, from the phase diagram to spectroscopy. We know that, although allowed by the symmetries, a term proportional to the topological charge, $S_{\theta}=i\theta Q_{\rm{top}}$, which would enhance those effects, is suppressed in practice. The natural instrument to study the effect this topological term would have on physics is lattice field theory. However, since this term is complex, it prevents the usual Monte Carlo methods to be effective and generates a sign problem.\ Here we will discuss the complex Langevin approach to the problem and show some results at imaginary and real $\theta$.

\section{Theory}
\label{Theory}

The Lagrangian we consider, after a Wick rotation, is the usual Yang-Mills action density plus the topological $\theta$-term,
\be 
\label{L_top}
 \mathcal{L} = \mathcal{L}_{\rm{YM}} - i \theta q(x) , \qquad\qquad 
 q(x)=\frac{g_0^2}{64\pi^2} F^a_{\mu\nu}(x)\tilde{F}^a_{\mu\nu}(x),
 \ee
where $q(x)$ is the topological charge density, $F^a_{\mu\nu}$ is the field strength tensor and $\tilde{F}^a_{\mu\nu}$ its Hodge dual tensor,
\be 
F^a_{\mu \nu}  = \partial_\mu A^a_\nu - \partial_\nu A^a_\mu + g \ f^{abc} A^b_{\mu}A^c_{\nu},
\qquad\qquad
\tilde{F}^a_{\mu\nu}= \dfrac{1}{2}\epsilon_{\mu\nu\rho\sigma}F^a_{\rho \sigma}.
\ee
The last term in the Lagrangian is a pseudo-scalar and, importantly, a total derivative which, when integrated over the four dimensional volume, corresponds to the \textit{topological charge},
\be
\label{Qt}
 \int d^4x\, q(x) = Q_{\rm{top}},  
 \ee
see e.g.\ Ref.\ \cite{Schafer:1996wv} for a review.
To carry out a non-perturbative study on the lattice, one needs to discretize the theory. Although it might seem that this procedure could destroy the topology of the configurations, it has been shown that topology emerges also on the lattice if the fields are smooth enough  \cite{Luscher:1981zq}.
  We adopt the naive discretisation of the topological term \cite{DiVecchia}  in Eq.\ \eqref{L_top},
\be 
\label{eq:qlat}
q_L(n) = - \dfrac{1}{2^4 \times 32 \pi^2} \sum_{\mu\nu\rho\sigma =\pm 1}^{\pm 4} \tilde{\epsilon}_{\mu\nu\rho\sigma} \Tr[\Pi_{\mu\nu}(n) \Pi_{\rho \sigma}(n)],
\ee 
where $\Pi_{\mu\nu}$ is the ordinary plaquette and the sum is over all directions back and forward, with $\tilde{\epsilon}_{-\mu\nu\rho\sigma}=-\epsilon_{\mu\nu\rho\sigma}$. While this expression reproduces the correct naive continuum limit, at finite lattice spacing it mixes with other operators and receives additive and multiplicative renormalization \cite{D'Elia:2003gr, DelDebbio:2002xa, Vicari:2008jw}
 \be
 q_L(n) \rightarrow a^4 Z_L(g^2) q(x) + O(a^6).
 \ee
To be able to measure the topological content of a configuration, in practice, one has eliminate the short-range quantum noise introduced by the lattice cutoff, by smoothing the configuration towards the semi-classical minima of the  Yang-Mills action. Only then integer values of the topological charge are recovered. This can be achieved using smearing, cooling, gradient flow, etc.\cite{Bonati:2014tqa,deForcrand:1997sq,Vicari:2008jw,Luscher:2010iy}. In Section \ref{WilsonFlow} we will discuss the application of the gradient flow to complexified configurations in \SL.

\section{Sign problem and complex Langevin dynamics}
\label{Sign_CL}

Standard Monte Carlo numerical methods use the Boltzman weight of a configuration to assign a probability  weight and to reconstruct a sample representative of the total space of configurations, in such a way that averages of observables can be computed over this sample. Such procedures fail when the action is not real and positive, since it is not easy to associate a probability to a configuration. In the case of the theta term some possibilities to get around this problem are to use reweighting or to  consider imaginary $\theta$ and then analytically continue the results to real $\theta$ \cite{Panagopoulos:2011rb,D'Elia:2012vv,Bonati:2014oqa}.

Our approach is to use complex Langevin dynamics in order to obtain the desired distribution as the asymptotic distribution of a stochastic process in the complexified configuration space, see Refs.\ \cite{Aarts:2013bla,Aarts:2013uxa} for recent reviews. The sign problem is evaded by this enlargement of the field space.
The stochastic process for links in \SU is described by
\be
\label{CL}
U(t+\epsilon) = R(U) U(t)\ , \qquad\qquad 
R(U) = \exp\left( -i\sum_a \lambda_a \left( \epsilon  D_a S[U] + \sqrt{\epsilon} \eta_a \right)\right),
\ee
where $t$ is the additional Langevin time in which the stochastic process takes place, $\epsilon$ is the Langevin time step, $\lambda_a$ are the generators of the group, and $\eta_a$ is stochastic gaussian noise that obeys 
\be
\langle \eta_a(t,x) \rangle = 0, \qquad\qquad 
\langle \eta_a(t,x)\eta_b(t',x')  \rangle = 2\delta_{ab}\delta(t-t') \delta(x-x').
\ee
As one can see, if the action $S$ is complex, Eq.\ \eqref{CL} takes the gauge links into the complex extension of the gauge group,
\be
U \in \rm{SL}(3,\mathbb{C}).
\ee
The probability distribution corresponding to the complexified stochastic process, formally a solution of the associated \textit{Fokker-Planck} equation, is positive and real even when the action $S$ is not. In that sense complex Langevin dynamics evades the sign problem and can, potentially, access e.g.\ the whole temperature-$\theta$ phase diagram. While for real Langevin dynamics one can analytically prove convergence of the \textit{Fokker-Planck} equation to the correct distribution ``a priori'' \cite{Damgaard:1987rr}, for complex Langevin such a proof relies on ``a posteriori'' checks of criteria of correctness. These criteria involve the distribution of observables to be \textit{localised} in the complex direction; when satisfied then a proof of correct convergence also exist for complex Langevin dynamic, provided that the drift is holomorphic \cite{Aarts:2009uq,Aarts:2011ax}. For non-holomorphic drifts additional care is required \cite{Mollgaard:2013qra}.

\section{Complex Langevin dynamics at imaginary $\theta$}

In order for complex Langevin dynamics to converge to the correct result, it is necessary to use gauge cooling \cite{Seiler:2012wz,Aarts:2013uxa,Bongiovanni:2013nxa}, i.e.\ to control the exploration of the complexified configuration space. In practice gauge cooling becomes insufficient when the gauge coupling $\beta \lesssim 5.7$ in the case of pure SU(3) gauge theories. This threshold depends on the presence of fermions \cite{Sexty:2013ica} but not on the lattice volume or, in this case, on the value of $\theta$. In fact, our way to test the efficiency of gauge cooling, and Langevin simulations in general, is to run the same procedure at $\theta=0$ or for imaginary $\theta$; even though the action is real, the Langevin approach will still attempt to explore the enlarged field space and gauge cooling is essential to constrain this (of course one could also occasionally re-unitarise  the links but this is not possible at real $\theta$). Hence it make sense to benchmark complex Langevin and gauge cooling at imaginary $\theta$.

\begin{figure}[t]
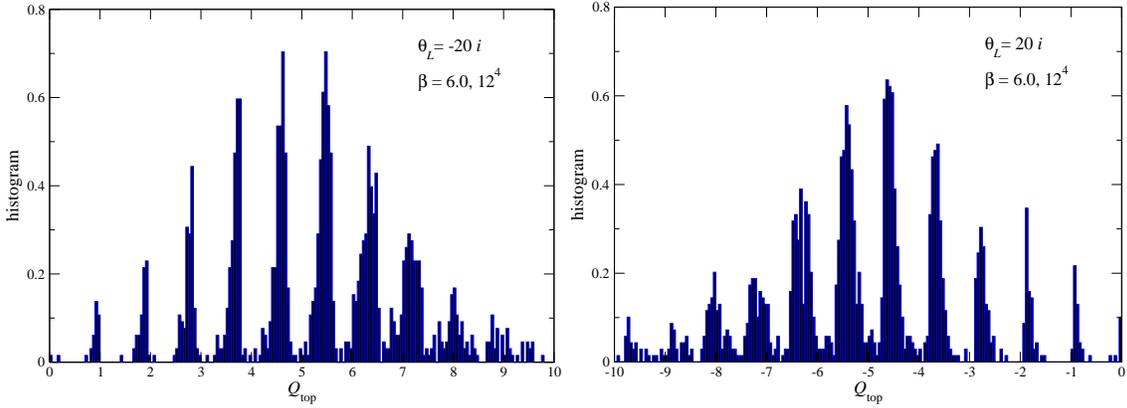
 
\begin{center}
\includegraphics[width=0.49\textwidth]{Figures/distribQ_th-20.eps} 
\includegraphics[width=0.49\textwidth]{Figures/distribQ_th20.eps} 
\end{center}
\caption{Distribution of $Q_{\rm top}$ at imaginary $\theta_L=-20i$ (left) and $\theta_L=+20i$ (right)  on a $12^4$ lattice at $\beta=6$, obtained by generating configurations using complex Langevin dynamics with gauge cooling and subsequently applying gradient flow to recover (close to) integer charges.
} 
\label{fig:distribQ} 
\end{figure}

In Fig.\ \ref{fig:distribQ} we show the result of such a test, at unrenormalised $\theta_L=\pm 20i$, and  $\beta=6$ on a $12^4$ lattice. Here configurations are generated using complex Langevin dynamics including gauge cooling to control the process. We subsequently use gradient flow to smoothing those and recover the topological content. We observe the expected response as the sign of $\theta_L$ is flipped. We have also made a comparison with  results obtained using the standard HMC algorithm and found agreement.

\section{Complex Langevin dynamics at real $\theta$}

Having verified our approach at imaginary $\theta$, we now present some results for real $\theta$. Writing the partition function as
\be
 Z(\theta_L) = \int DU \ e^{-S_{\rm YM}} e^{i\theta_L Q} = e^{-\Omega f(\theta_L)},
 \ee
 where $\Omega$ is the lattice four-volume and $f(\theta_L)$ the free energy density, it is straightforward to predict the 
behaviour of the topological charge $Q$ for real and imaginary $\theta_L$ as
\begin{align}
\label{eqQ}
 -i\langle Q \rangle_{\theta_L} &\, = - \dfrac{\partial\ln Z}{\partial\theta_L} = \Omega \chi_L  \theta_L\left( 1 + 2 b_2\theta_L^2 + 3 b_4\theta_L^4 + \ldots\right), \\
 -\langle Q \rangle_{\theta_I} &\, = \;\;\;\,\dfrac{\partial\ln Z}{\partial\theta_I}  = \Omega \chi_L \theta_I\left( 1- 2 b_2 \theta_I^2 + 3 b_4 \theta_I^4 + \ldots\right),
 \end{align}
where $\theta_I = -i\theta_L$ and $\chi_L$ is the (unrenormalised) topological susceptibility.
Here and below $\theta_L$ always denotes the unrenormalised $\theta$ parameter in the action and $Q$ denotes the topological charge obtained directly from the lattice simulations, see Eq.\ (\ref{eq:qlat}), i.e.\ without smoothening of the gauge fields.
Hence the coefficients $b_i$ in the expansion are also unrenormalised.

\begin{figure}[t]
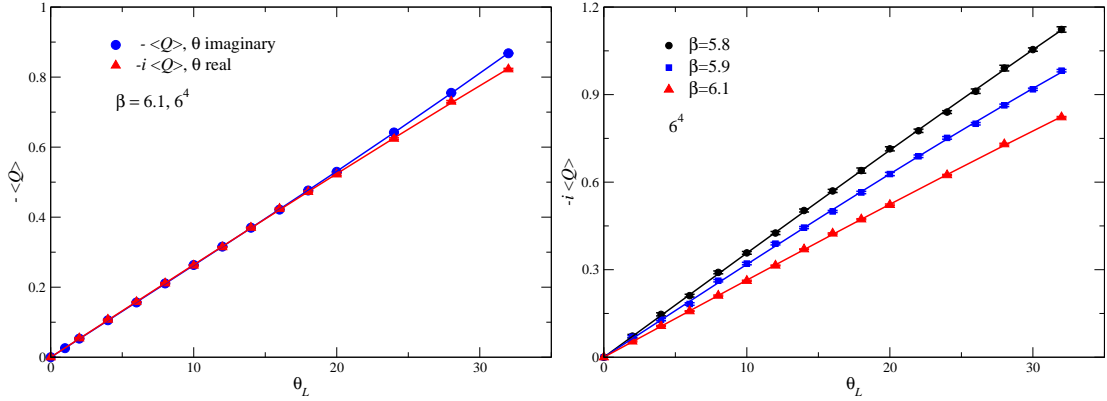
 
\begin{center}
\includegraphics[width=0.48\textwidth]{Figures/QReQI_b6.1-GA.eps} 
\includegraphics[width=0.48\textwidth]{Figures/QvsTh_variousBeta-GA.eps}
\end{center}
\caption{Left: Dependence of $\bra Q\ket$ on the unrenormalised $\theta_L$, both real and imaginary, at $\beta=6.1$. Right: Dependence on real $\theta_L$ at $\beta=5.8, 5.9, 6.1$.
 } 
\label{fig:Q(th)} 
\end{figure}

\begin{figure}[t] 
\begin{center}
\includegraphics[width=0.58\textwidth]{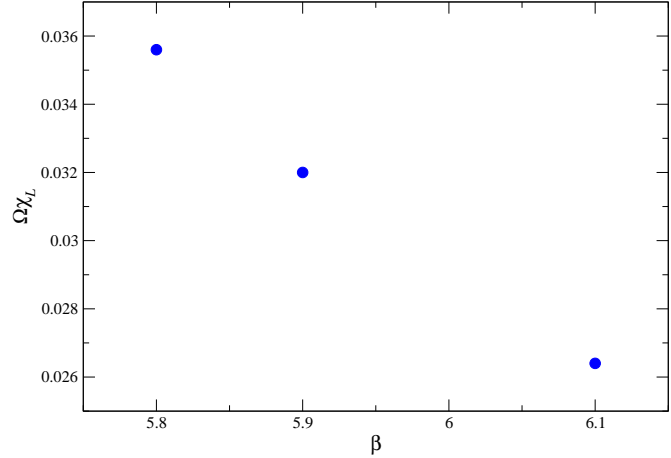}
\end{center}
\caption{Dependence of the linear term, $\Omega\chi_L$, on $\beta$. 
} 
\label{fig:Chi_top} 
\end{figure}

In Fig.\ \ref{fig:Q(th)} (left) we show the dependence of $\langle Q \rangle$ on both imaginary and real $\theta_L$, at $\beta=6.1$ on a $6^4$ lattice. We observe an almost perfect linear increase, with a deviation between the real and imaginary lines only visible at the larger $\theta_L$ values. Independent fits to both data sets yields agreement with the analytical prediction, i.e.\ a common linear term and small third-order term with opposite sign. Explicitly,
\be
y(\theta_L) = \Omega\chi_L \theta_L (1 \pm 2 b_2 \theta^2_L), 
\qquad\qquad
\Omega\chi_L = 0.026, \qquad\qquad b_2\sim10^{-5}.
\ee
Repeating this for other $\beta$ values we find the $\theta_L$ dependence shown in Fig.\ \ref{fig:Q(th)} (right) and a dependence of the linear coefficient $\Omega\chi_L$ on $\beta$ as in Fig.\  \ref{fig:Chi_top}.
As expected, the susceptibility decreases with increasing coupling (decreasing lattice spacing or increasing temperature).

\section{Gradient flow on \SL}
\label{WilsonFlow}

As specified in Sec.\ \ref{Theory}, all the topological operators on the lattice need to be renormalised due to short range quantum fluctuations that couple with the operator \eqref{Qt}. The non-perturbative way to do that involves cooling the configurations towards the local minima of the Yang-Mills action.  In Sec.\ \ref{Sign_CL} we mentioned, however, that the gauge group complex Langevin dynamics takes place in is not SU(3) but instead \SL. The latter is a non-compact group with unstable classical directions in the noncompact direction, and therefore it is not obvious whether cooling/smoothening will have the same effect as in SU(3).

\begin{figure}[t] 
\begin{center}
\includegraphics[width=0.58\textwidth]{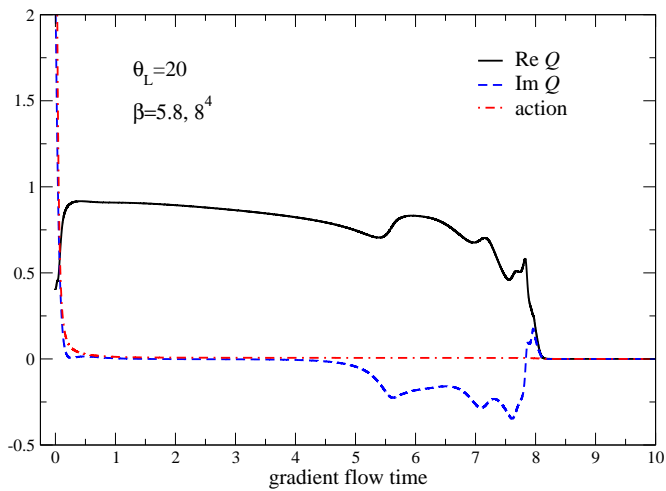}
\end{center}
\caption{Typical behaviour of the real and imaginary parts of the lattice topological charge $Q$ and the action under the gradient flow for an \SL\ configuration.  } 
\label{fig:WilsonFlow} 
\end{figure}

Since the gradient flow is just Langevin dynamics without noise (i.e.\ ``classical evolution''), it is straightforward to implement the procedure in \SL. 
We have done various tests but have not yet gained a satisfactory understanding. A typical example is given in Fig.\ \ref{fig:WilsonFlow}, where the gradient flow evolution of the charge $Q$ and the action $S$ are given. We observe a plateau in the real part of $Q$, but with values that average to zero in the ensemble. This is not surprising, since the expectation value of $Q$ should be purely imaginary, see Eq.\ (\ref{eqQ}). However, the imaginary part of $Q$ flows to zero quickly (which is not surprising either). Perhaps, this conundrum reflects the fact that complex Langevin dynamics works on the level of expectation values of holomorphic observables and not on the level of individual \SL\ configurations, as manipulating those violates holomorphicity.

\section{Summary}

We have shown that we have  good control of the lattice theory for $\theta$ both real and imaginary. The criteria for the correctness of complex Langevin dynamics are satisfied for $\beta \gtrsim 5.8$ and the observables behave in a sensible way around $\theta=0$. We demonstrated that the behaviour of the lattice topological charge follows nicely the prediction based on analytic continuation from imaginary $\theta$. As an application we studied the dependence of the (unrenormalised) topological susceptibility $\chi_L$ on $\beta$.
We conclude that complex Langevin dynamics correctly simulates the lattice theory, in terms of the bare parameters and lattice operators.

The next step is to express our findings in terms of the renormalised $\theta=Z_L\theta_L$, where the renormalisation factor can be determined at $\theta=0$. 
The challenge ahead is to find a way to obtain information on renormalised topological observables from the Langevin dynamics in \SL.

\section*{Acknowledgments}
We thank Ion-Olimpiu Stamatescu for many useful discussions and Massimo d'Elia and Francesco Negro for correspondence. We are grateful for the computing resources made available by HPC Wales and by STFC through DiRAC computing facilities. This work is supported by STFC, the Royal Society, the Wolfson Foundation and the Leverhulme Trust.

\end{document}